\documentclass[letterpaper, 11pt]{article}

\usepackage{mathtools}
\usepackage{array}
\usepackage{enumitem}
\usepackage{amsmath,amsfonts,amssymb,bm}
\usepackage{titlesec}
\usepackage{pgfgantt,rotating}
\usepackage{float}
\usepackage{graphicx}
\usepackage{sidecap}
\usepackage{wrapfig}
\usepackage{bold-extra}
\usepackage{authblk}
\usepackage{appendix}
\usepackage{makecell}
\usepackage{dirtytalk}
\usepackage{placeins}
\usepackage{subfigure}
\usepackage[export]{adjustbox}
\usepackage{cite}

\usepackage[margin=1in]{geometry}

\newcommand{\beq}{\begin{equation}}
\newcommand{\eeq}{\end{equation}}
\newcommand{\bgqar}{\begin{eqnarray}}
\newcommand{\enqar}{\end{eqnarray}}
\newcommand{\bgqarn}{\begin{eqnarray*}}
\newcommand{\enqarn}{\end{eqnarray*}}
\newcommand{\bgary}{\begin{array}}
\newcommand{\enary}{\end{array}}

\graphicspath{{Figures/}}

\title{A Graph Theoretic Approach in Combination With Dynamic Mode Decomposition With Control (DMDc) to Analyze Battery Degradation}

\author{   Khalid Mahmud Labib\affil{1}, 
	Saad Waheed\affil{1}, 
	Bakhtiar Nafis\affil{1}, 
	Shabbir Ahmed\affil{1}}

\affil{Dynamical Systems and Signals Lab (DSSL)\\ Department of Mechanical Engineering\\South Dakota State University, Brookings, SD }

\begin{document}

\maketitle


\begin{abstract}

Accurate monitoring of lithium-ion battery (LIB) degradation is essential, yet it remains challenging due to the complex, nonlinear, and time-varying nature of electrochemical aging processes. Conventional equivalent circuit models (ECMs) provide simplified representations of battery behavior using fixed electrical components, but they cannot capture evolving internal degradation mechanisms and structural changes over time. In this study, a data-driven framework is developed by integrating dynamic mode decomposition with control (DMDc) with graph-theoretic analysis to characterize battery degradation from operational data alone. The mode matrix ($\mathbf{\phi}$) obtained from DMDc is transformed into a weighted adjacency matrix, enabling the representation of battery dynamics as an evolving network of interacting states. Graph-based measures, including connectivity and a modularity (proxy), are then used to quantify structural changes in the system across degradation stages. The results show a clear transition from a highly connected and coherent network in the healthy state to a progressively weaker and more fragmented structure as degradation advances, accompanied by increasing heterogeneity. This work demonstrates that graph-theoretic representations can effectively capture the evolving dynamics of battery degradation and provide interpretable insights into system-level aging behavior.

\end{abstract}



\newpage\pagebreak 

\tableofcontents 

\section{Introduction}

The global demand for clean and sustainable energy solutions has driven advancements in battery technology for energy storage systems. LIBs have gained significant attention in this industry because of their high energy density, low self-discharge rate, and environmental benefits~\cite{Li2025}. However, repeated charging and discharging lead to irreversible electrochemical processes that increase cell impedance and reduce capacity. This degradation results from mechanisms such as solid electrolyte interphase (SEI) formation, lithium-ion loss, and electrolyte decomposition~\cite{Xu2025,Safitri2026}. Once degradation surpasses a certain threshold, potential safety hazards such as electrolyte leakage, localized short circuits, and thermal runaway may occur. Therefore, accurate estimation of battery degradation and knowledge about the state of health (SOH) is essential.

Traditional battery monitoring approaches can be broadly categorized into physics-based and data-driven frameworks. Physics-based models, particularly electrochemical models, are derived from first principles and aim to capture the underlying mechanisms governing battery behavior. These models provide high-fidelity representations by describing ion transport, charge transfer, and diffusion processes through coupled partial differential equations (PDEs). However, despite their strong physical interpretability, their practical implementation requires extensive parameter identification, state estimation, and model calibration using experimental data~\cite{Deng2017}, making them computationally intensive and challenging to deploy in real-time applications.

In contrast, data-driven approaches have gained increasing attention due to their computational efficiency and ease of implementation. Among these, equivalent circuit models (ECMs) offer a simplified representation of battery dynamics by modeling the system using electrical components such as resistors and capacitors, with parameters including ohmic resistance ($R_{0}$), polarization resistance ($R_{p}$), and capacitance ($C_{p}$)~\cite{Lai2018}. These models rely explicitly on measured signals such as voltage, current, and temperature to learn input–output relationships. While ECMs are well-suited for real-time applications, their parameters are typically tuned to fit experimental data and depend heavily on the assumed circuit topology. Consequently, they lack direct physical interpretability with respect to underlying electrochemical processes and are limited in their ability to capture the structural evolution associated with battery degradation.

Recently, machine learning (ML)-based data-driven approaches have widely been adopted for modeling battery degradation due to their ability to capture complex, nonlinear relationships directly from operational data. Regression-based methods such as Gaussian process regression (GPR) combined with extended Kalman filtering (EKF), and kernel support vector machines (SVMs) optimized via metaheuristic algorithms, have demonstrated strong performance in state-of-health (SOH) estimation across varying operating conditions \cite{Pang2022,Liu2024}. Sequence models, including long short-term memory (LSTM) networks, effectively capture temporal degradation trends from historical capacity trajectories, while ensemble learning techniques such as XGBoost further enhance predictive accuracy through feature integration \cite{Zhang2023,Zhao2023}.

In practical battery systems, degradation evolves through complex interactions driven by thermal, electrical, and mechanical coupling, leading to heterogeneous and spatially correlated aging behavior. In this work, we aim to employ graph-theoretic approach for characterizing battery degradation, which provides a natural framework for representing interconnected systems. To form a network, we employed the mode matrix obtained from dynamic mode decomposition as a graph \(G(V,E)\), where nodes represent system states and edges represent coupling dynamics between them. Graph theory has been widely applied to complex dynamical systems across multiple domains. In chemistry, molecular structures are represented as bond graphs to explore reaction pathways and predict physicochemical properties through quantitative structure–property relationships (QSPR)~\cite{Ramos2023,Fraz2025}. In biological systems, graph-based models have been used to identify mutation clusters and analyze protein stability under varying environmental conditions~\cite{Ryslik2014,Miotto2024,Prabantu2025}. Similarly, in fluid dynamics, graph representations of interacting vortices enable reduced-order modeling and efficient analysis of flow behavior~\cite{Wang2025}. These applications demonstrate the effectiveness of graph theory in capturing the structural and dynamical evolution of complex systems with time. Previously, SOH and remaining useful life (RUL) have been predicted using graph convolutional networks based on voltage or feature sequences \cite{He2024,Wei2023}. However, no prior work was found which utilizes graph theoretic approach to understand battery degradation.

To address these limitations, this study proposes a graph-theoretic framework for modeling battery degradation by leveraging the mode matrix obtained from dynamic mode decomposition with control (DMDc) algorithm applied to hybrid pulse power characterization (HPPC) data \cite{Labib2026}. DMDc is a data-driven system identification method that estimates a linear dynamical operator $\mathbf{A}$ governing the evolution of measured states such as the terminal voltage under known inputs such as the applied current. Through eigen-decomposition, this operator yields a mode matrix $\boldsymbol{\Phi}$, whose entries encode the amplitude and phase relationships among dynamical modes. This matrix is reinterpreted here as a weighted adjacency matrix, wherein nodes correspond to delay-embedded state coordinates or latent system states inferred from the measured voltage response, and edge weights quantify the strength and directionality of dynamic coupling among those states which is expected to shift systematically as the cell ages.

Based on this learned adjacency structure, graph-theoretic metrics such as connectivity, modularity, and their temporal evolution over charge-discharge cycles are computed to characterize the battery degradation. These metrics provide a systematic means to quantify how the underlying interaction structure evolves as degradation progresses. By constructing the graph directly from data without imposing a predefined topology, the proposed approach captures emergent coupling patterns associated with aging. This framework thus establishes a principled connection between data-driven dynamical systems theory (DMDc) and graph analysis, offering new insights into the structural evolution of battery degradation.

The remainder of this paper is organized as follows. Section 2 presents the theoretical framework, including an overview of graph theory and DMDc. Section 3 describes the data generation process and experimental setup. Section 4 presents the experimental results and discussion, demonstrating the proposed method's application to battery degradation data. Section 5 discusses the current approach's limitations and outlines directions for future work. Finally, Section 6 summarizes the key findings and concludes the paper.

\section{Methodology}

\begin{figure}[]
\includegraphics[width=1\linewidth]{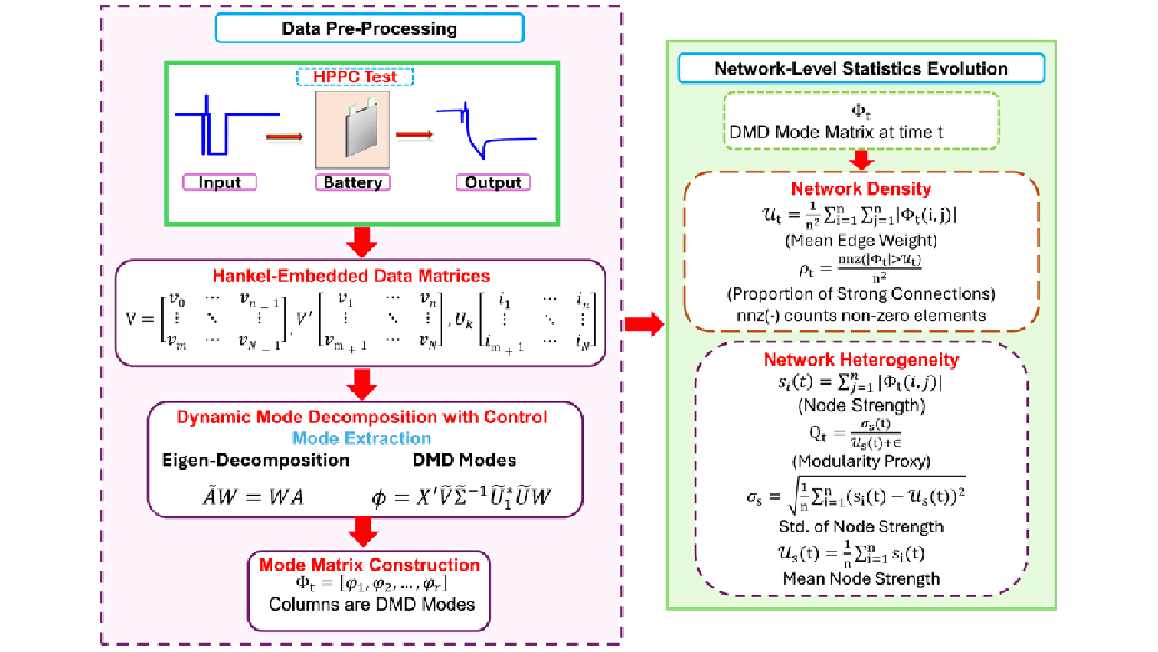}
\caption{Schematic representation of the DMDc enhanced graph theoretic approach used in this study to analyze battery degradation}
\label{fig:graph}
\end{figure}

This section presents the theoretical framework for modeling battery degradation using graph theory and DMDc, which can be visualized from Figure~\ref{fig:graph}. By interpreting the DMDc-derived system matrix as a weighted adjacency matrix, we construct an evolving network of cell interactions and compute graph metrics that reveal degradation-driven changes in system organization. These metrics provide a system-level, topology-free representation of how degradation propagates across the battery pack over charge-discharge cycles.

\subsection{Dynamic Mode Decomposition with Control}

In this study, the DMDc framework is employed to identify a linear state-space representation of the battery dynamics from data. The formulation follows the methodology presented in \cite{Labib2026}, where the battery system is described as

\begin{equation}\label{eq:dmdc_data}
    \mathbf{x_{k+1}} \approx \mathbf{A}\,\mathbf{x_k} + \mathbf{B}\,\mathbf{u_k},
\end{equation}

To obtain a reduced-order representation, singular value decomposition (SVD) is applied to both $\mathbf{X}'$ and $\mathbf{\Omega}$, and the projection is carried out onto a low-dimensional subspace as described in \cite{Labib2026}. This leads to the reduced-order system matrix $\tilde{\mathbf{A}}$, given by

\begin{equation}
\tilde{\mathbf{A}} \approx \hat{\mathbf{U}}^{*} \mathbf{X}' \tilde{\mathbf{V}} \tilde{\mathbf{\Sigma}}^{-1} \tilde{\mathbf{U}}_{x}^{*} \hat{\mathbf{U}},
\end{equation}

The spectral properties of the system are then obtained by solving the eigenvalue problem

\begin{equation}
\tilde{\mathbf{A}} \mathbf{W} = \mathbf{W} \mathbf{\Lambda},
\end{equation}
where $\mathbf{\Lambda}$ contains the eigenvalues and $\mathbf{W}$ contains the corresponding eigenvectors.

Finally, the DMD modes are reconstructed as
\begin{equation}
\mathbf{\phi} = \mathbf{X}' \tilde{\mathbf{V}} \tilde{\mathbf{\Sigma}}^{-1} \tilde{\mathbf{U}}_{x}^{*} \hat{\mathbf{U}} \mathbf{W}.
\end{equation}

These modes provide a compact representation of the system dynamics and form the basis for subsequent network-based analysis.

\subsection{Graph Theory Framework for Battery Degradation}

We represent the system dynamics using a weighted graph $\mathcal{G}_t = (\mathcal{V}, \mathcal{E}_t, \mathbf{\phi}_t)$, where each $\mathbf{\phi}_t \in \mathbb{R}^{n \times n}$ serves as the adjacency matrix at cycle $t$, and where each vertex $v_i$ corresponds to a delay-embedded state coordinate inferred from the measured voltage response. The entry, $\mathcal{E}_t(i,j)$ quantifies the spatiotemporal coupling dynamics from state $j$ to state $i$, thereby encoding the underlying dynamics of the system. In this framework, $\phi_t$ is obtained directly from the DMDc algorithm, which identifies the state-transition operator from measured voltage responses and input currents, as illustrated in Figure~\ref{fig:graph}.

\subsubsection{Connectivity (Network Density)}

\begin{figure}[]
    \centering

    \subfigure{
        \includegraphics[width=0.48\linewidth]{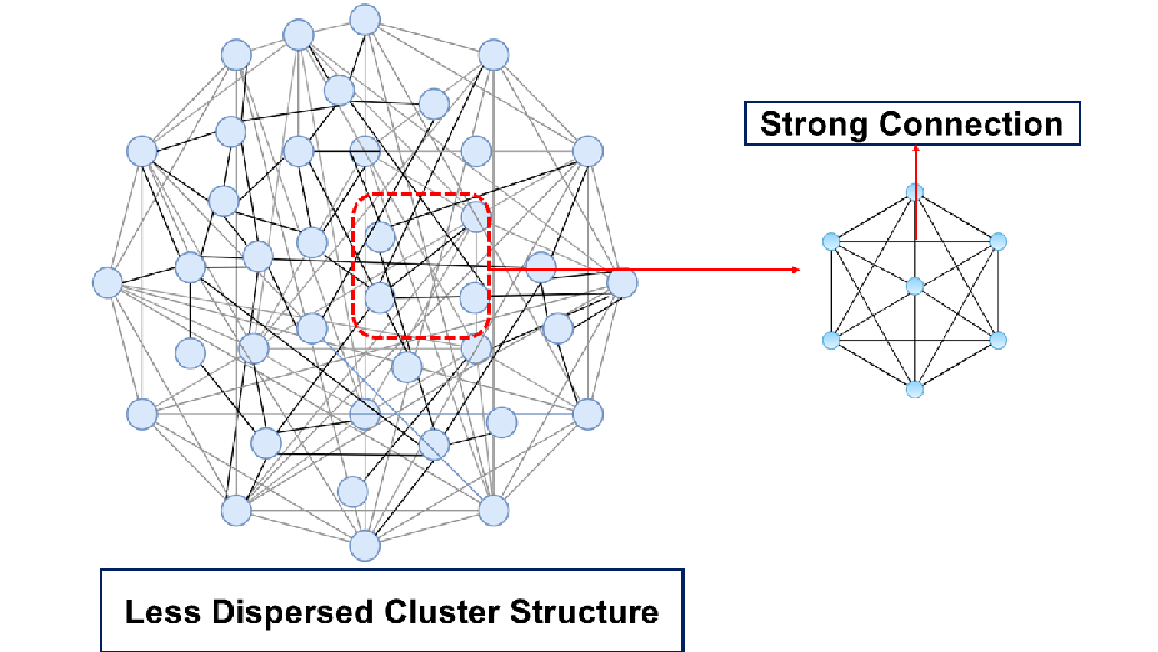}
        \label{fig:Healthy_network}
    }
    \hfill
    \subfigure{
        \includegraphics[width=0.48\linewidth]{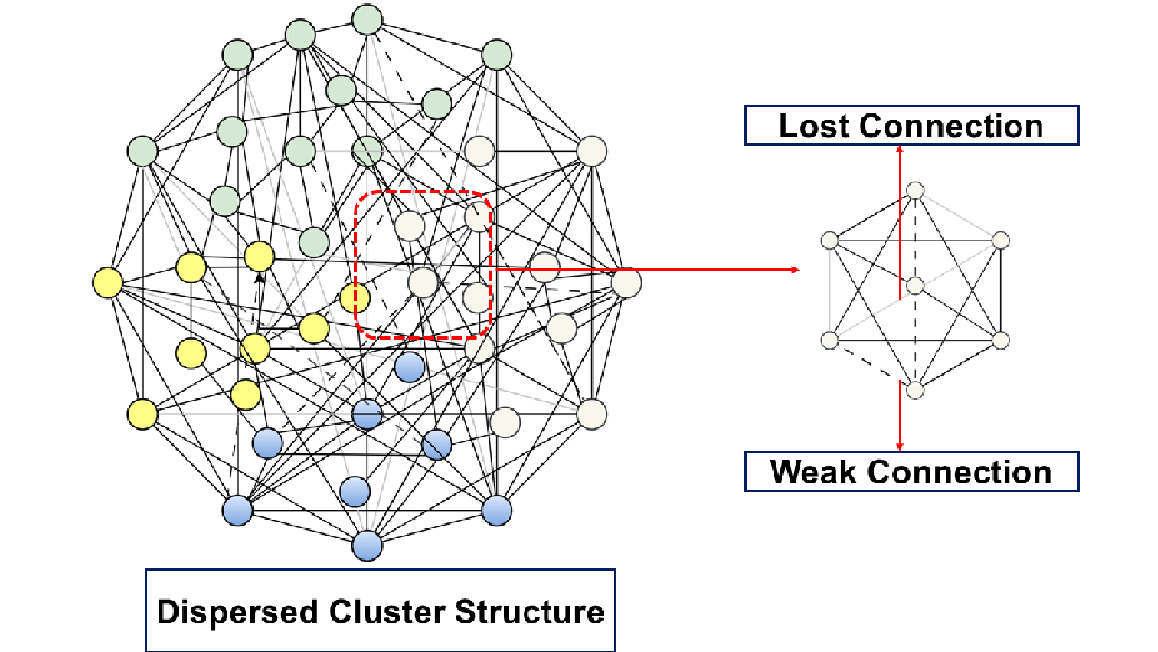}
        \label{fig:Degraded_network}
    }

    \caption{Comparison of network topology in (a) healthy and (b) degraded states. In the healthy condition, connections are predominantly strong and uniformly distributed, while in the degraded state, the emergence of weakened (dashed) and severed (light gray) links reflects the loss of connectivity and increased structural fragmentation.}
    \label{fig:Graph_Network}
\end{figure}

To characterize the structural evolution of the system, the DMD mode matrix $\mathbf{\phi}_t$ is interpreted as a weighted adjacency matrix, where each entry represents the strength of interaction between nodes. In this framework, the nodes correspond to delay-embedded state variables, and edge weights encode the inferred dynamic coupling between these states obtained from the DMDc framework. The overall idea of connectivity with battery degradation is shown in Figure \ref{fig:Graph_Network}

In line with graph-theoretic network analysis, connectivity is evaluated using a thresholded representation of the weighted network, where only sufficiently strong interactions are retained. An adaptive threshold is defined based on the mean magnitude of the matrix entries:
\begin{equation}
\mu_t = \frac{1}{n^2} \sum_{i=1}^{n} \sum_{j=1}^{n} |\phi_t(i,j)|.
\end{equation}
This choice allows the threshold to adapt to variations in the overall interaction strength across different system states, avoiding the need for an arbitrary fixed value.

A binary adjacency matrix $\mathbf{A}_t \in \{0,1\}^{n \times n}$ is then constructed as
\begin{equation}
A_t(i,j) =
\begin{cases}
1, & |\phi_t(i,j)| > \mu_t, \\
0, & \text{otherwise},
\end{cases}
\end{equation}
thereby retaining only the relatively strong interactions while discarding weaker connections.

Next, the total number of active edges in the binary adjacency matrix is computed as
\begin{equation}
N_t = \sum_{i=1}^{n} \sum_{j=1}^{n} A_t(i,j),
\end{equation}
where $N_t$ denotes the number of retained connections, i.e., the total number of entries equal to one in $\mathbf{A}_t$.

The connectivity is then defined as the fraction of active edges relative to the total number of possible connections:
\begin{equation}
\rho_t = \frac{N_t}{n^2}.
\end{equation}

This formulation yields a normalized, dimensionless measure of connectivity that reflects the proportion of significant interactions within the system. As such, it enables consistent comparison across different operating conditions and degradation stages while preserving a clear physical interpretation in terms of interaction density.

\subsubsection{Modularity Proxy (Network Heterogeneity)}

Modularity is a measure of the strength of community structure in a network, defined as the difference between the actual number of edges within communities and the expected number of such edges in a random network with the same degree distribution.

In the present framework, explicit community detection is not performed. Instead, the heterogeneity of the network is characterized using a modularity-inspired metric based on the distribution of node strengths. The mode matrix $\mathbf{\phi}_t$ is interpreted as a weighted adjacency matrix, where each entry represents the strength of interaction between nodes. Variations in these interaction strengths across nodes reflect the degree of structural heterogeneity in the system.

To evaluate this, the magnitude of the adjacency matrix is first considered:
\begin{equation}
\mathbf{\Phi}_t \leftarrow |\mathbf{\phi}_t|.
\end{equation}

The node strength of the $i$-th node is then computed as the sum of all connections associated with that node:
\begin{equation}
s_i(t)=\sum_{j=1}^{n}\Phi_t(i,j), \qquad i=1,2,\dots,n.
\end{equation}

The mean node strength is defined as
\begin{equation}
\mu_s(t)=\frac{1}{n}\sum_{i=1}^{n}s_i(t),
\end{equation}

and the standard deviation of node strength is computed as
\begin{equation}
\sigma_s(t)=\sqrt{\frac{1}{n}\sum_{i=1}^{n}\left(s_i(t)-\mu_s(t)\right)^2}.
\end{equation}

A modularity proxy is then defined as the normalized dispersion of node strengths:
\begin{equation}
Q_t^{\text{proxy}}=\frac{\sigma_s(t)}{\mu_s(t)+\epsilon},
\end{equation}
where $\epsilon$ is a small positive constant introduced to avoid division by zero.

This formulation captures the heterogeneity of the network structure. A lower value of $Q_t^{\text{proxy}}$ indicates that node strengths are distributed more uniformly, corresponding to a less dispersed and more homogeneous network. In contrast, higher values indicate increased variability in node strengths, reflecting a more heterogeneous structure with localized or unevenly distributed interactions.
\section{Data Generation}

A 30 Ah lithium-ion battery was tested using a hybrid pulse power characterization (HPPC) protocol to study its dynamic behavior within a voltage range of 2.5 V to 4.2 V. The test consisted of a sequence of impulse charge, discharge, and rest steps with controlled current pulses, repeated to capture different states of charge. After performing the HPPC test on the healthy battery, the cell was subjected to continuous charge–discharge cycling to induce degradation, up to 360 cycles. The same HPPC test was conducted periodically every 20 cycles to monitor changes in the voltage response with given input current sequence and evaluate its degradation over time. The collected responses were then used to construct snapshot matrices and to form the $\mathbf{\phi}$ matrices following the DMDc methodology described earlier. More details about the experiment are provided in \cite{Labib2026}.

\section{Results and Discussions}

This section presents the results of applying the proposed graph-theoretic framework to battery degradation data. First, the spatial characteristics of the DMDc modes are examined to illustrate how degradation alters the spatiotemporal coherence of the system. Next, the adjacency matrices derived from the DMD mode matrices are analyzed, followed by a quantitative assessment of network connectivity and modularity evolution across different degradation stages. These metrics provide insight into the structural reorganization of the battery network as aging progresses.

\subsection{Spatial Characterization of DMDc Modes}
The DMDc modes provide a combined spatiotemporal representation of the system dynamics, where the magnitude and phase together describe the spatial participation and temporal evolution of the underlying coherent structures. The magnitude of each mode reflects the strength of contribution from different spatial locations, while the phase represents the relative temporal lag across the domain. Together, these quantities enable a comprehensive interpretation of how the system dynamics evolve under degradation.

\begin{figure}[H]
    \centering

    \subfigure{
        \includegraphics[width=0.48\linewidth]{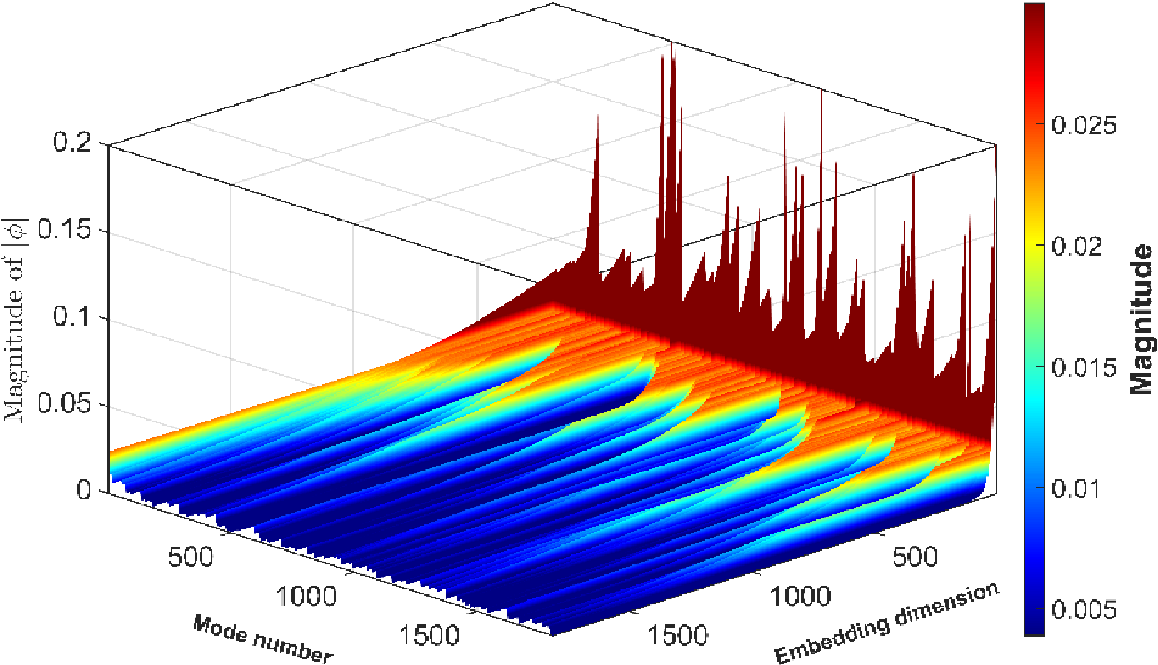}
        \label{fig:abs_h}
    }
    \hfill
    \subfigure{
        \includegraphics[width=0.48\linewidth]{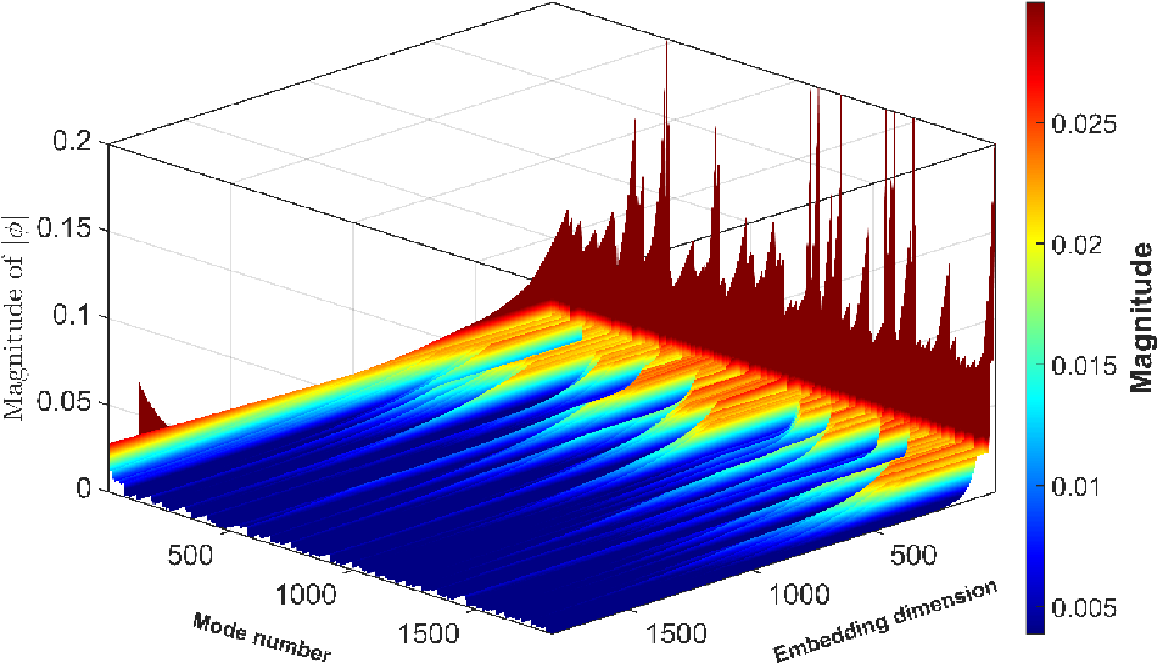}
        \label{fig:abs_360}
    }

    \caption{Spatial distribution of DMD mode magnitude for (a) the healthy state and (b) the degraded state. The magnitude represents the spatial intensity of the coherent structure, indicating regions of strong and weak modal influence.}
    \label{fig:absolute}
\end{figure}
Figure~\ref{fig:absolute} presents the spatial distribution of modal magnitude for both healthy and degraded conditions. A clear contrast between the two battery states is observed in the organization and distribution of magnitude across the surface. In the region where mode number is lower than 500 (first 500 modes), the degraded state exhibits a localized increase in magnitude relative to the healthy case (distinct spikes emerge), particularly in the region where embedding dimension is greater than 1500. This indicates that certain modes which were previously weak begin to contribute more significantly. Along the embedding dimension direction, the healthy case is characterized by a smooth and continuous variation in magnitude, forming well-defined ridge-like structures. However, the degraded state shows noticeable irregularities and distortions in these patterns, reflecting a loss of structural coherence. Specially, in the intermediate region where mode number is greater than 700 and less than 1100 ($700< \text{mode number}< 1100$), the overall magnitude is elevated in the degraded state, however, the distribution spike magnitude becomes less uniform, suggesting that energy is redistributed rather than uniformly increased. In the region where mode number is greater than 1500 and embedding dimension is less than 300 ($\text{mode number} > 1500$) and ($\text{embedding dimension} < 300$), a modest but consistent increase in magnitude is observed, indicating that previously inactive modes begin to exhibit weak contributions to overall system. Moreover, the degraded state is marked by more frequent and irregular spikes across the surface, particularly for $\text{embedding dimension} = 1800$, implying that the energy is no longer concentrated in a few dominant modes but is more dispersed. Overall, the surface evolves from a structured and smoothly varying pattern to a more irregular and scattered distribution, reflecting a redistribution of modal energy.

\begin{figure}[H]
    \centering

    \subfigure{
        \includegraphics[width=0.48\linewidth]{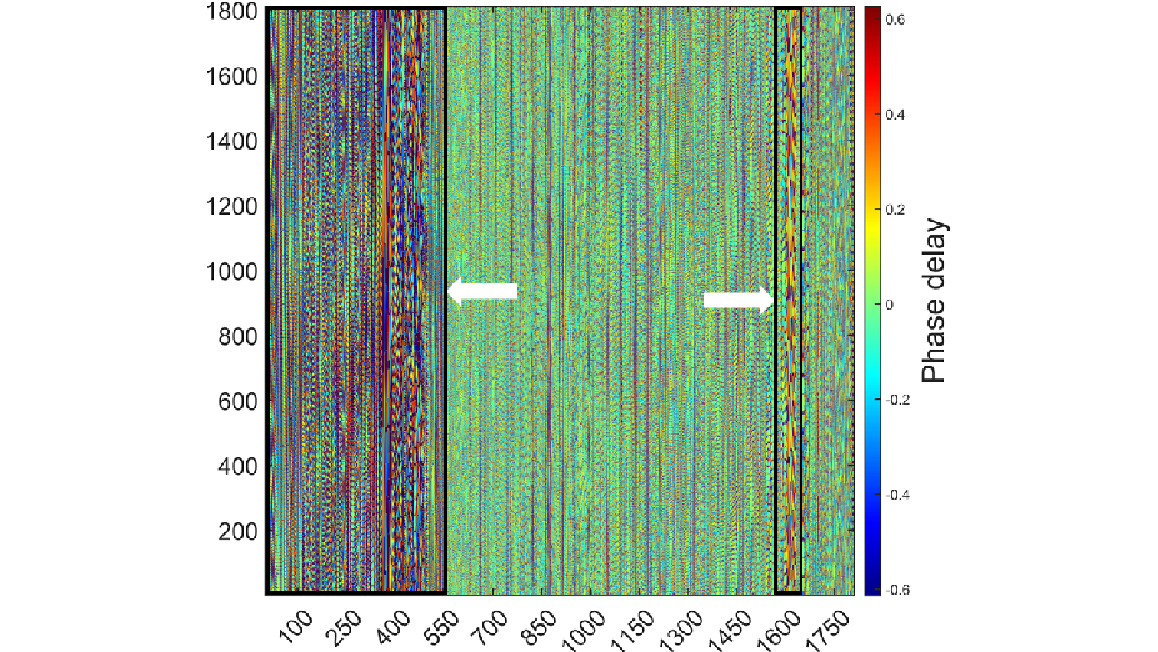}
        \label{fig:phase_h}
    }
    \hfill
    \subfigure{
        \includegraphics[width=0.48\linewidth]{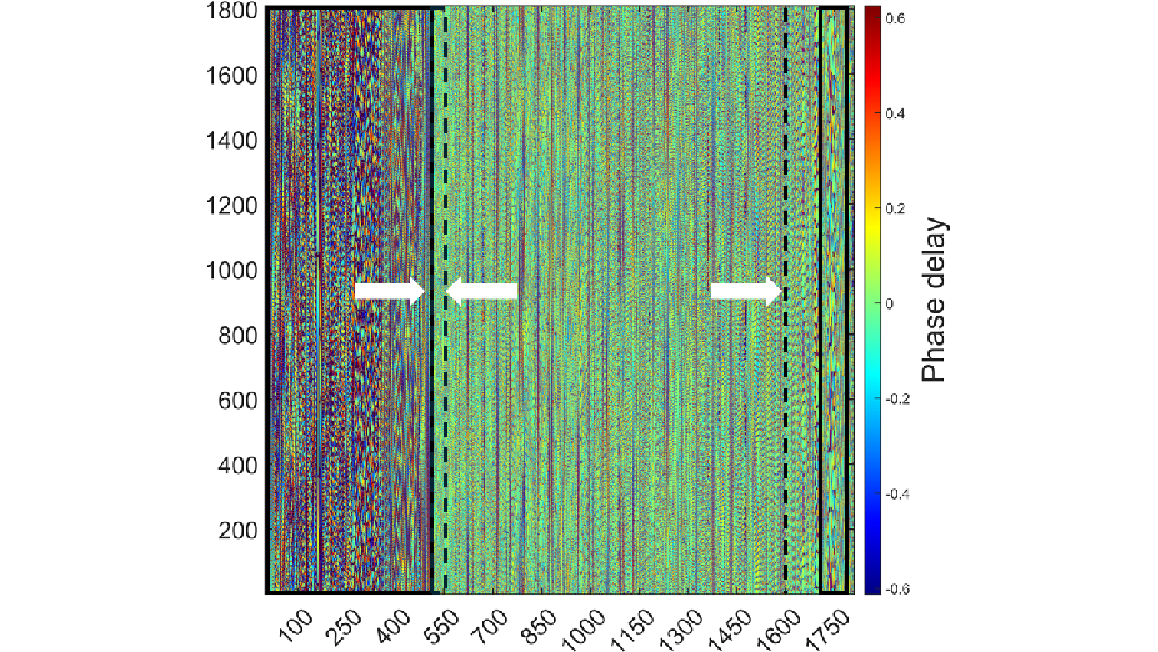}
        \label{fig:phase_360}
    }

    \caption{Spatial distribution of DMD mode phase for (a) the healthy state and (b) the degraded state. The phase represents the local temporal change of each spatial location, indicating how different regions of the structure oscillate relative to one another.}
    \label{fig:phase}
\end{figure}

The phase delay distribution in Figure~\ref{fig:phase} highlights the temporal behavior of the system. As shown in Figure~\ref{fig:phase}(a), the initial region in the healthy state shows a wider coverage of higher magnitude phase extending up to approximately 550. This indicates that a larger set of modes contributes actively and in a coordinated manner. In the degraded state (Figure~\ref{fig:phase}(b)), this region reduces to about 500 (white arrow), suggesting that some of these modes lose their influence. The degraded case also shows more irregular phase spikes, indicating that the remaining modes act in a less coordinated way. This reflects a reduction in the consistency of modal contributions across the domain. In the later region, the healthy case shows localized and structured phase activity near 1600. In contrast, the degraded state shows similar activity shifted to higher indices, around 1750 (white arrow).

Overall, while the healthy state maintains a relatively coherent spatiotemporal structure with smoother magnitude transitions and more organized phase behavior in energetically significant regions, the degraded state exhibits both increased spatial fragmentation and reduced temporal coherence. These observations suggest that degradation primarily affects the consistency of the underlying dynamics, with phase variations providing a more sensitive indicator of structural changes than magnitude alone.

\subsection{Adjacency matrices}

Different aging mechanisms, such as SEI formation, side reactions, and structural changes, contribute to battery degradation. Developing a comprehensive understanding of these processes requires accounting for their strong path dependency and the complex interaction structure.In this context, the DMD modes, represented by the matrix $\mathbf{\phi}$, provide a data-driven characterization of the underlying spatiotemporal behavior at different degradation stages. Since the elements of $\mathbf{\phi}$ encode the spatial distribution and interaction of dynamic features,as discussed in earlier section, across the system, the matrix can be interpreted as a weighted adjacency representation. This interpretation enables the application of network-based measures, such as connectivity and modularity, to quantify the structural organization of the system. This perspective helps quantify the heterogeneity of the network structure and examine how connectivity evolves relative to the initial stage of battery degradation.

\subsection{Network Connectivity}

In this framework, an adaptive threshold defined by the mean value of each $\mathbf{[\phi_{ij}]}$ matrix is employed to convert the weighted matrix into a binary adjacency matrix. Connections exceeding this mean value are assigned a value of one, while all others are set to zero. This approach eliminates the need for arbitrary fixed thresholds and ensures that the network representation adapts to the underlying system dynamics. Connectivity is then quantified as the fraction of retained edges relative to the total number of possible edges. This normalization ensures that the metric is dimensionless and enables consistent comparison across different conditions.
\begin{figure}[H]
\includegraphics[width=\linewidth]{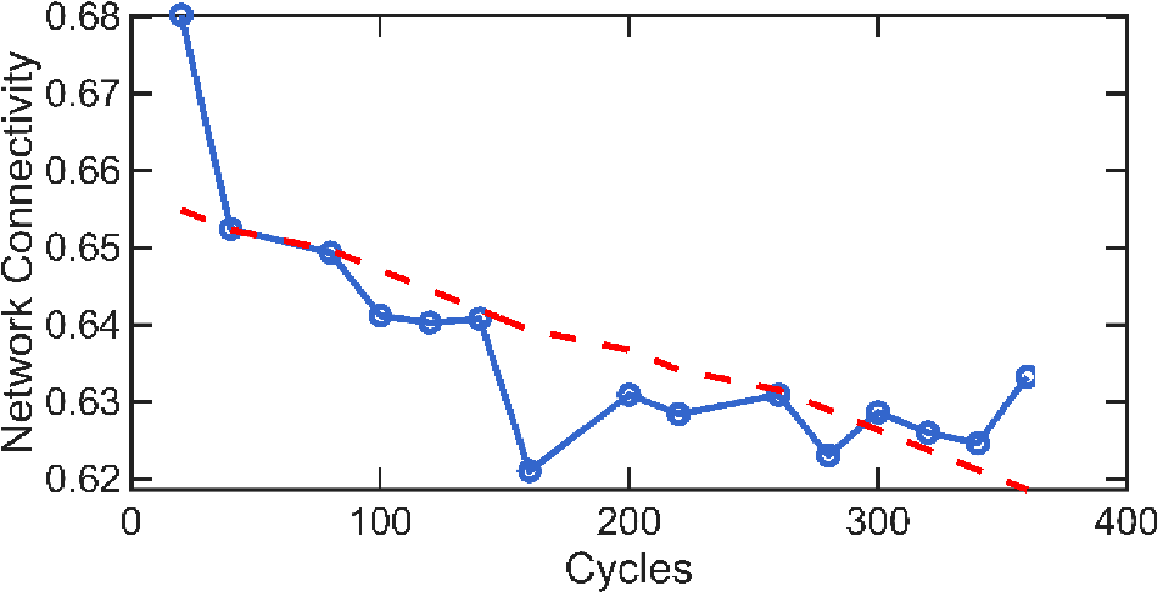}
\caption{Evolution of connectivity between nodes over different degradation stages.}
\label{fig:connectivity}
\end{figure}

From Figure~\ref{fig:connectivity}, it can be observed that at the initial stage, corresponding to the healthy state of the battery, the connectivity attains its highest value of 0.68. As the battery undergoes degradation, the connectivity gradually decreases, reaching 0.653 by the 40th cycle, then 0.623 by 340th cycle. This continuous reduction indicates a progressive loss of strong connections within the network as the battery ages. In this context, higher connectivity reflects a more stable and well-connected network structure, whereas lower connectivity signifies a weakening of interactions, indicating a transition toward a more fragile network as degradation progresses.

\subsection{Modularity}

Modularity is a community-based metric that requires partitioning the network into groups and comparing intra-community connectivity against a random-network baseline. In the present study, however, the adopted formulation does not recover this classical definition. Instead, it quantifies the dispersion of node strengths and therefore functions as a surrogate measure of structural heterogeneity rather than a direct measure of community modularity.

\begin{figure}[H]
\includegraphics[width=\linewidth]{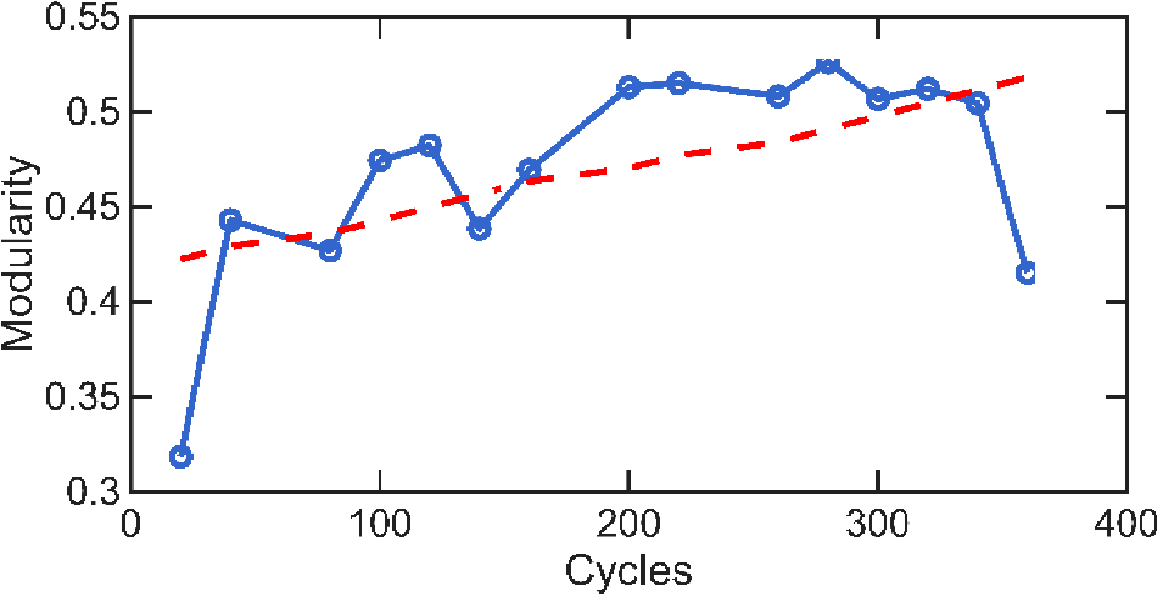}
\caption{Comparison of Modularity over different degradation stages.}
\label{fig:modularity}
\end{figure}

From Figure \ref{fig:modularity}, it can be seen that, the modularity values lie within the range $\mathbf{[0.3,0.7]}$ \cite{gunduz2017community}, which is expected for natural communities. In the healthy state, the modularity is approximately 0.3, implying a comparatively uniform and less dispersed distribution of clusters. With increasing degradation, however, various ageing mechanisms progressively reshape the network structure. This is reflected in the rise of modularity, which indicates that the interaction patterns become more heterogeneous and that the cluster structure evolves toward a more dispersed and nonuniform configuration.



\section{Limitations and Future Recommendations}

The proposed framework relies on the interpretation of DMD modes, where the magnitude represents spatial participation and the phase represents relative temporal evolution across the domain. While this provides a comprehensive spatiotemporal description of the system, several limitations should be noted. The results are influenced by modeling choices such as rank truncation, embedding dimension, and sampling, all of which can affect the stability and interpretability of the extracted modes.

Additionally, the phase angle of the DMD modes becomes unreliable in regions where the modal magnitude is small. In such cases, phase variations may appear random and may not correspond to physically meaningful temporal behavior. The presence of noise and high-frequency dynamics may further introduce artificial spikes in the magnitude distribution and irregular phase variations, which can influence the derived network measures.

Moreover, the DMD mode matrix $\mathbf{\phi}$ is treated as a weighted adjacency matrix; however, its entries do not correspond to direct physical connections. Instead, they represent modal interactions in the spatiotemporal domain, making the resulting network an abstract representation of the system dynamics. The connectivity metric is also defined using a mean-based threshold, which, although simple and adaptive, remains heuristic and not physically unique. Since threshold-based measures are sensitive to the choice of threshold, alternative selections may lead to variations in the inferred network structure. Consequently, the observed reduction in connectivity may be partially influenced by the thresholding strategy rather than solely reflecting intrinsic system dynamics.

Future research should focus on developing formulations that more rigorously link DMD-derived matrices to physically interpretable network structures, potentially through Koopman operator theory or correlation-based representations. Advanced DMD techniques, such as optimized DMD, sparsity-promoting DMD, or total least-squares DMD, can be employed to reduce the influence of noise and improve robustness.

Furthermore, the proposed framework can be extended to multi-modal datasets, higher-dimensional systems, and real-time monitoring scenarios, including applications to other energy storage technologies and complex structural systems. Finally, graph-based machine learning approaches, such as Graph Neural Networks (GNNs), offer a promising direction by utilizing DMD-derived adjacency matrices to learn complex spatiotemporal patterns, enabling improved feature extraction, damage classification, and predictive modeling of system degradation.

\section{Conclusions}

In this study, graph-theoretic approaches, specifically connectivity and modularity, are employed to analyze the evolution of battery dynamics in the spatiotemporal domain as degradation mechanisms progressively develop. DMDc is applied to obtain the $\mathbf{\phi}$ matrices at different stages of degradation. These matrices are interpreted as weighted adjacency matrices, enabling the use of network-based measures to quantify properties such as connectivity and structural heterogeneity in the evolving system.

For connectivity, an adaptive thresholding approach is utilized to ensure that the network representation reflects the underlying dynamics at each stage. At the healthy state, most connections exceed the mean threshold value, resulting in a highly connected and stable network structure. As degradation progresses, many interactions weaken and fall below the threshold, leading to a reduction in connectivity. In contrast, the modularity proxy exhibits an increasing trend with degradation. At the healthy state, the modularity value is relatively low (approximately 0.3), indicating a more uniform and less dispersed network structure. As the system degrades, the network becomes increasingly heterogeneous, with interactions becoming unevenly distributed, resulting in more dispersed cluster formations.

The novelty of this work lies in establishing a connection between DMDc-based system identification and graph-theoretic analysis. The results demonstrate the potential of network-based approaches to provide new insights into the evolving dynamics associated with degradation mechanisms. Overall, this study opens new avenues for understanding degradation processes not only in LIBs but also in a broader class of energy storage systems.
\section*{Acknowledgments}
This work is supported by National Science Foundation (NSF), United States, Award Number: 2501703 and CRG grant from South Dakota Board of Regents (SDBOR).


\nocite{*}

\bibliographystyle{asmeconf}  
\bibliography{references}

@article{Li2025,
  author  = {Li, Q. and Song, R. and Wei, Y.},
  title   = {A review of state-of-health estimation for lithium-ion battery packs},
  journal = {Journal of Energy Storage},
  year    = {2025},
  volume  = {118},
  pages   = {116078}
}

@article{Xu2025,
  author  = {Xu, W. and others},
  title   = {Investigation of lithium-ion battery degradation by corrected differential voltage analysis based on reference electrode},
  journal = {Applied Energy},
  year    = {2025},
  volume  = {389},
  pages   = {125735}
}

@article{Safitri2026,
  author  = {Safitri, M. and Adji, T.B. and Cahyadi, A.I.},
  title   = {Enhanced adaptive LSTM framework for robust early-life prediction of lithium-ion battery degradation},
  journal = {Results in Engineering},
  year    = {2026},
  volume  = {29},
  pages   = {109664}
}

@article{Deng2017,
  author  = {Deng, Z. and others},
  title   = {Implementation of reduced-order physics-based model and multi-parameters identification strategy for lithium-ion battery},
  journal = {Energy},
  year    = {2017},
  volume  = {138},
  pages   = {509--519}
}

@article{Lai2018,
  author  = {Lai, X. and Zheng, Y. and Sun, T.},
  title   = {A comparative study of different equivalent circuit models for estimating state-of-charge of lithium-ion batteries},
  journal = {Electrochimica Acta},
  year    = {2018},
  volume  = {259},
  pages   = {566--577}
}

@article{Pang2022,
  author  = {Pang, B. and Chen, L. and Dong, Z.},
  title   = {Data-Driven Degradation Modeling and SOH Prediction of Li-Ion Batteries},
  journal = {Energies},
  year    = {2022},
  volume  = {15},
  number  = {15},
  pages   = {5580}
}

@article{Liu2024,
  author  = {Liu, S. and others},
  title   = {State-of-health estimation of lithium-ion batteries using a kernel support vector machine tuned by a new nonlinear gray wolf algorithm},
  journal = {Journal of Energy Storage},
  year    = {2024},
  volume  = {102},
  pages   = {114052}
}

@article{Zhang2023,
  author  = {Zhang, L. and others},
  title   = {Accurate Prediction Approach of SOH for Lithium-Ion Batteries Based on LSTM Method},
  journal = {Batteries},
  year    = {2023},
  volume  = {9},
  number  = {3},
  pages   = {177}
}

@article{Zhao2023,
  author  = {Zhao, Y. and others},
  title   = {Data-driven lithium-ion battery degradation evaluation under overcharge cycling conditions},
  journal = {IEEE Transactions on Power Electronics},
  year    = {2023},
  volume  = {38},
  number  = {8},
  pages   = {10138--10150}
}

@article{Ramos2023,
  author  = {Ramos-Sánchez, P. and Harvey, J.N. and Gámez, J.A.},
  title   = {An automated method for graph-based chemical space exploration and transition state finding},
  journal = {Journal of Computational Chemistry},
  year    = {2023},
  volume  = {44},
  number  = {1},
  pages   = {27--42}
}

@article{Fraz2025,
  author  = {Fraz, H.M. and others},
  title   = {Predictive modeling of copper iodide properties using graph-theoretical descriptors},
  journal = {Scientific Reports},
  year    = {2025},
  volume  = {15},
  number  = {1},
  pages   = {39878}
}

@article{Ryslik2014,
  author  = {Ryslik, G.A. and others},
  title   = {A graph theoretic approach to utilizing protein structure to identify non-random somatic mutations},
  journal = {BMC Bioinformatics},
  year    = {2014},
  volume  = {15},
  number  = {1},
  pages   = {86}
}

@article{Miotto2024,
  author  = {Miotto, M. and others},
  title   = {Osmolyte-induced protein stability changes explained by graph theory},
  journal = {Computational and Structural Biotechnology Journal},
  year    = {2024},
  volume  = {23},
  pages   = {4077--4087}
}

@article{Prabantu2025,
  author  = {Prabantu, V.M. and others},
  title   = {The alteration of structural network upon transient association between proteins studied using graph theory},
  journal = {Proteins: Structure, Function, and Bioinformatics},
  year    = {2025},
  volume  = {93},
  number  = {1},
  pages   = {217--225}
}

@article{Wang2025,
  author  = {Wang, Z. and others},
  title   = {Development and application of a fluid mechanics analysis framework based on complex network theory},
  journal = {Computer Methods in Applied Mechanics and Engineering},
  year    = {2025},
  volume  = {435},
  pages   = {117677}
}

@article{Wu2024,
  author  = {Wu, T. and others},
  title   = {Research on Two-Level Equalization Strategy of Lithium-Ion Battery Based on Graph Theory},
  journal = {Journal of Electrochemical Energy Conversion and Storage},
  year    = {2024},
  volume  = {21},
  number  = {2},
  pages   = {021013}
}

@article{Wu2023,
  author  = {Wu, T. and Wang, Y. and Zhang, J.},
  title   = {Research on multilayer fast equalization strategy of Li-ion battery based on adaptive neural fuzzy inference system},
  journal = {Journal of Energy Storage},
  year    = {2023},
  volume  = {67},
  pages   = {107574}
}

@article{Dao2011,
  author  = {Dao, T.-S. and McPhee, J.},
  title   = {Dynamic modeling of electrochemical systems using linear graph theory},
  journal = {Journal of Power Sources},
  year    = {2011},
  volume  = {196},
  number  = {23},
  pages   = {10442--10454}
}

@article{Kartha2021,
  author  = {Kartha, T.R. and Mallik, B.S.},
  title   = {Molecular dynamics and emerging network graphs of interactions in dinitrile-based Li-ion battery electrolytes},
  journal = {The Journal of Physical Chemistry B},
  year    = {2021},
  volume  = {125},
  number  = {26},
  pages   = {7231--7240}
}

@article{He2024,
  author  = {He, Y. and others},
  title   = {State-of-health estimation for fast-charging lithium-ion batteries based on a short charge curve using graph convolutional and long short-term memory networks},
  journal = {Journal of Energy Chemistry},
  year    = {2024},
  volume  = {98},
  pages   = {1--11}
}

@article{Wei2023,
  author  = {Wei, Y. and Wu, D.},
  title   = {Prediction of state of health and remaining useful life of lithium-ion battery using graph convolutional network with dual attention mechanisms},
  journal = {Reliability Engineering \& System Safety},
  year    = {2023},
  volume  = {230},
  pages   = {108947}
}

@misc{Labib2026,
  author = {Labib, K.M. and Ahmed, S.},
  title  = {Modeling of Non-linear Dynamics of Lithium-ion Batteries via Delay-Embedded Dynamic Mode Decomposition},
  year   = {2026},
  note   = {arXiv preprint arXiv:2601.22403}
}

@inproceedings{gunduz2017community,
  title={Community detection in social media network with maximum modularity using girvan-newman algorithm},
  author={G{\"u}nd{\"u}z, Ali Fatih and Karado{\u{g}}an, Ahmet},
  booktitle={SETSCI-Conference Proceedings},
  volume={1},
  pages={222--225},
  year={2017},
  organization={SETSCI-Conference Proceedings}
}

\end{document}